\documentclass[preprint2]{aastex}
\usepackage{rotating}

\shortauthors{Crepp et al. 2010}

\begin{document}
\setcounter{secnumdepth}{2}
\title{Speckle Suppression with the Project 1640 Integral Field Spectrograph}

\author{Justin R. Crepp\altaffilmark{1}, Laurent Pueyo\altaffilmark{2,7,8}, Douglas Brenner\altaffilmark{3}, Ben R. Oppenheimer\altaffilmark{3}, Neil Zimmerman\altaffilmark{3,4}, Sasha Hinkley\altaffilmark{1}, Ian Parry\altaffilmark{5}, David King\altaffilmark{5}, Gautam Vasisht\altaffilmark{2}, Charles Beichman\altaffilmark{6,1,2}, Lynne Hillenbrand\altaffilmark{1}, Richard Dekany\altaffilmark{1}, Mike Shao\altaffilmark{2}, Rick Burruss\altaffilmark{2}, Lewis C. Roberts, Jr.\altaffilmark{2}, Antonin Bouchez\altaffilmark{1}, Jenny Roberts\altaffilmark{2}, R\'emi Soummer\altaffilmark{7}}
\email{jcrepp@astro.caltech.edu}

\altaffiltext{1}{California Institute of Technology, 1200 E. California Blvd., Pasadena, CA 91125} 
\altaffiltext{2}{Jet Propulsion Laboratory, 4800 Oak Grove Drive, Pasadena, CA 91109}
\altaffiltext{3}{American Museum of Natural History, Central Park West at 79th Street, New York, NY 10024} 
\altaffiltext{4}{Columbia University, 550 West 120th Street, New York, NY 10027}
\altaffiltext{5}{University of Cambridge, Madingley Rd., Cambridge, CB3 OHA, UK}
\altaffiltext{6}{NASA Exoplanet Science Institute, 770 S. Wilson Avenue, Pasadena, CA 911225}
\altaffiltext{7}{Space Telescope Science Institute, 3700 San Martin Drive, Baltimore, MD 21218}
\altaffiltext{8}{Johns Hopkins University, 3400 N. Charles Street, Baltimore, MD 21218}

\begin{abstract}
Project 1640 is a high-contrast imaging instrument recently commissioned at Palomar observatory. A combination of a coronagraph with an integral field spectrograph (IFS), Project 1640 is designed to detect and characterize extrasolar planets, brown dwarfs, and circumstellar material orbiting nearby stars. In this paper, we present our data processing techniques for improving upon instrument raw sensitivity via the removal of quasi-static speckles. Our approach utilizes the chromatic image diversity provided by the IFS in combination with the locally-optimized combination of images (LOCI) algorithm to suppress the intensity of residual contaminating light in close angular proximity to target stars. We describe the Project 1640 speckle suppression pipeline (PSSP) and demonstrate the ability to detect companions with brightness comparable to and below that of initial speckle intensities using on-sky commissioning data. Our preliminary results indicate that suppression factors of at least one order of magnitude are consistently possible, reaching $5\sigma$ contrast levels of $2.1\times10^{-5}$ at $1\arcsec$ in the H-band in 20 minutes of on-source integration time when non-common-path errors are reasonably well-calibrated. These results suggest that near-infrared contrast levels of order $\approx10^{-7}$ at subarcsecond separations will soon be possible for Project 1640 and similarly designed instruments that receive a diffraction-limited beam corrected by adaptive optics (AO) systems employing deformable mirrors with high actuator-density. 
\end{abstract}
\keywords{techniques: spectroscopic, high angular resolution -- instrumentation: adaptive optics}  

\section{Introduction}\label{sec:intro}
Wavefront errors created by imperfections in the optical components (mirrors, lenses, beam splitters, ... etc.) of a high-contrast imaging instrument manifest as a complex pattern of quasi-static intensity variations, or so-called ``speckles", in the focal plane at the science camera. Nominally bright compared to the signal of a substellar companion or debris disk, speckle intensities must first be minimized by hardware, usually with an adaptive optics (AO) system, and then further suppressed through data processing. Since individual speckles have the same intrinsic width as the instrument point-spread-function (PSF), this latter step is crucial for distinguishing between stellar artifacts and true companions. 

Speckle suppression is accomplished through differential imaging by exploiting differences between the properties of residual starlight that reaches the detector and any incoherent radiation arriving from an off-axis source. To preserve companion flux, some form of ``diversity" must be introduced into the system, whereby a sequence of images is recorded such that individual frames are trivially different from one another. Once measured or calibrated, the image diversity is reversed using software to remove the speckles and reveal previously hidden companions.  

Image diversity can be achieved in a number of ways. Examples include modulation of companion position with respect to speckles through instrument rotation or field rotation \citep{krist_07,marois_08,marois_10}, modulation of input polarization state \citep{potter_03,oppenheimer_08,hinkley_09}, and modulation of the actual target being observed \citep{serabyn_10,crepp_10,mawet_09}. The level to which residual starlight is removed depends on the modulation timescale compared to the speckle decorrelation timescale. Quasi-static speckles\footnote{Other families of speckles related to the atmosphere and adaptive optics system can decorrelate on a much faster timescale \citep{macintosh_05}.} generally decorrelate in tens or hundreds of seconds as the result of changes in ambient conditions and optical system alignment (e.g., \cite{hinkley_07}). Ideally, science images and PSF reference images are recorded simultaneously to minimize the influence of speckle pattern evolution.

An alternative approach to discriminate between companions and starlight is to generate diversity through the wavelength dependence of speckles. In this case, modulation of companion position with respect to speckles is achieved naturally through diffraction by acquiring images simultaneously in multiple different spectral channels. Chromatic differential imaging not only has the benefit of ``freezing" the speckle pattern in time, but it also yields the spectrum of any companions, thus enabling study of their chemical composition and bulk physical properties. First proposed by \cite{sparks_ford_02}, concurrent search and characterization may be accomplished in practice using an integral-field spectrograph (IFS). 

Compared to other speckle suppression techniques \citep{absil_mawet_09,oppenheimer_hinkley_09}, chromatic differential imaging has a high duty-cycle efficiency and can help maximize instrument scientific return by: accommodating a relatively wide bandpass\footnote{A special case of chromatic differential imaging, called simultaneous differential imaging (SDI), operates over a narrow bandpass and has shown promise for high-contrast applications involving methanated sources as demonstrated by the NICI campaign \citep{biller_10,liu_10}.}; allowing for the inter-changeable usage of science images and PSF references; maintaining an inner-working-angle that is independent of target coordinates relative to the observatory; avoiding smearing of the companion or dust disk PSF during an exposure; and obviating the need to observe a nearby PSF calibration star with similar brightness and colors. Chromatic differential imaging may also be implemented in combination with other techniques.

AO-assisted integral field spectroscopy has been used previously to study the companions orbiting GQ Lupi \citep{mcelwain_07}, AB Doradus \citep{thatte_07}, AB Pictoris \citep{bonnefoy_10}, and recently the outer-planets of HR 8799 \citep{janson_10,bowler_10}. In this paper, we present the first on-sky experiments that combine an IFS with a coronagraph \citep{hinkley_10_instrument} and speckle suppression using chromatic differential imaging \citep{sparks_ford_02}. To this, we also implement the locally optimized combination of images (LOCI) algorithm which improves the signal-to-noise ratio of detections \citep{lafreniere_07}. 

Project 1640 (hereafter, P1640) is a ground-based high-contrast imaging instrument that was recently installed and tested on the Hale 200-inch telescope at Palomar. The hardware incorporates a near-infrared coronagraph and IFS and will soon receive a corrected beam from the PALM-3000 ``extreme" AO system \citep{hinkley_10_instrument,bouchez_09,soummer_05}. P1640 has made several discoveries to date, including Alcor B and Zeta Virginis B \citep{zimmerman_10,hinkley_10}. Though faint compared to their host stars, these companions have masses of $\approx0.25M_{\odot}$ and $\approx0.17M_{\odot}$ respectively, and were noticed in raw data. 

In the following, we describe our technique for detecting companions having brightness and angular separation that places them beneath the noise floor set by speckles prior to data processing. This paper is accompanied by a companion paper by Pueyo et al. 2011 that discusses our method for subsequently extracting their spectrum. Combined, our results demonstrate the two principal utilities of using an IFS for high-contrast observations. These techniques are relevant to forth-coming instruments with similar designs, including the Gemini Planet Imager \citep{macintosh_GPI_06}, VLT SPHERE \citep{beuzit_06}, and Subaru HiCIAO \citep{mcelwain_10aas}, which will also use an IFS for chromatic differential imaging and companion characterization. 

\section{Data Pipeline Description}\label{sec:pipeline}
\subsection{IFS Raw Data}
The P1640 IFS records 40,000 different spectra across the instrument field of view with every exposure. Each spectrum corresponds to an individual element of a 200$\:$x$\:$200 lenslet array and is mapped to a different location in the image plane. Each lens in the array has a pitch of $\approx$19.2 mas as projected onto the sky. The P1640 spectral resolution is 23 channels across the J and H bands, spanning wavelengths $1.10 \leq \lambda \leq 1.75$ $\mu$m \citep{hinkley_10_instrument}. Raw spectra are converted into a data cube consisting of two spatial coordinates and one color coordinate using the procedure described in detail in Zimmerman et al. 2010b (submitted). Each extracted cube is flat-fielded, background-subtracted, and wavelength-calibrated based on observations of spectral standards. The {\it P1640 speckle suppression pipeline} (PSSP) starts with extracted and spectrally calibrated data cubes. 

\begin{figure*}[!t]
\begin{center}
\includegraphics[height=1.75in]{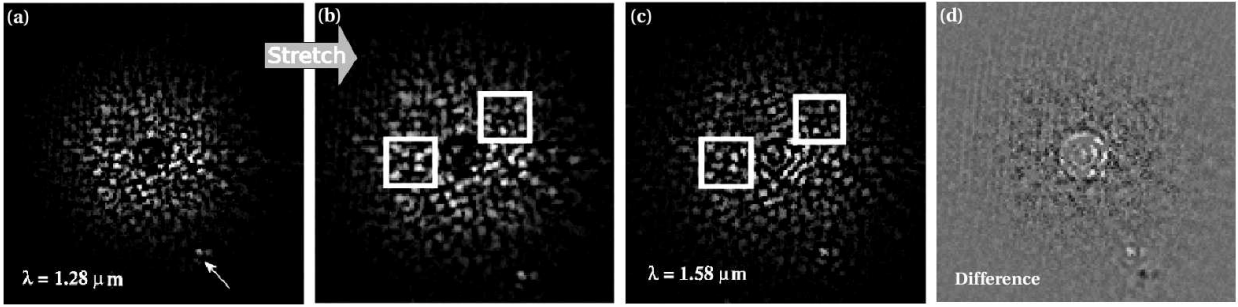} 
\caption{Occulted images of the star Alcor in individual IFS channels illustrating differential imaging with chromatic speckle diversity. Each image has a width of $3\arcsec \times\: 3\arcsec$. (a) Single channel in the J-band after applying a high-pass Fourier filter but prior to image stretching. The position of Alcor B is marked with an arrow. (b) Same image radially stretched to optimally match the speckle pattern for a longer wavelength channel. (c) Single channel in the H-band. The speckle structure is highly correlated between the middle panels (see interior of white boxes), enabling precision PSF alignment and subtraction. (d) Differenced image displayed on a linear intensity scale. The signal of a real companion is incoherent with starlight and survives post-processing because its position is independent of wavelength. Non-differenced images are displayed on a square-root intensity scale for comparison of the speckle patterns.} 
\end{center}\label{fig:speckles_space}
\end{figure*} 

\subsection{PSSP Code}
The PSSP is a custom program written in Matlab. This language was selected because it is flexible and contains a number of useful, fast, and robust built-in functions. Moreover, Matlab allows for a trivial transition from running scripts in serial on one CPU core to running the same scripts in parallel on multiple cores, facilitating handling of the large number of images and calculations per image. To enhance reduction speed, we use the Bluedot super-computing cluster at the NASA Exoplanet Science Institute (NExScI) which incorporates 16 nodes with 8 cores (2.5 GHz) per node. This feature permits pseudo real-time reduction while at the observatory, which enables follow-up observations of promising targets on subsequent nights within the same run to improve the fidelity of detections, spectra, and astrometry, and possibly take advantage of improved seeing conditions.

\subsection{Technique}
At its core, the PSSP involves precision stretching and shrinking of individual images to utilize the spatial dependence of the speckle pattern on wavelength. This technique takes advantage of the fact that the position of a true companion is constant in each channel while speckles move radially outwards from the star as a function of wavelength. For instance, the dependence of speckle position in the focal plane scales linearly with wavelength for phase aberrations created in the pupil plane. This form of diversity lends itself to differential imaging ($\S$\ref{sec:intro}). Stretching and shrinking images separated by several wavelength channels simultaneously aligns speckles and mis-aligns companions (Fig. 1). Upon subtraction, the highly correlated speckles are removed and companions survive, dramatically improving the effective detection sensitivity ($\S$\ref{sec:results}).

The multitude of available reference images provided by the IFS naturally lends itself to application of the locally optimized combination of images (LOCI) algorithm \citep{lafreniere_loci}. Once images are optimally stretched for a given wavelength channel, we apply LOCI to construct an optimal reference, including both color and time, by weighting individual images with coefficients based on a least-squares matching of local speckle intensities via matrix inversion. LOCI has been used with other speckle suppression techniques and provides higher signal-to-noise ratio detections compared to PSF subtraction involving a single reference image \citep{marois_08,mawet_09,lafreniere_09,crepp_10}. Details of the algorithm are discussed below.  

\subsection{Program Outline}
An outline of the PSSP procedure is shown in Figure 2. We begin by retrieving extracted data cubes output by the Zimmerman et al. 2010b raw IFS conversion program. Data headers are then culled for pertinent information and the code is preconditioned by the user. This latter step involves specifying which images to use based on data quality; searching for astrometric fiducial spots\footnote{A grid of wires placed in the pupil plane diffracts starlight into four bright spots centered on the star and may be used for astrometric purposes \citep{siv_06,zimmerman_10}.}; generating FOV, high-pass Fourier filter, and other alignment masks; setting basic parameters related to image cleaning and the LOCI algorithm; inserting artificial companions to assess the fidelity of the reduction; and initializing the number of Matlab ``workers" for parallel processing. Both occulted and unocculted images are loaded in order to perform relative photometry and calculate sensitivity in terms of contrast.  

\begin{figure*}[!t]
\begin{center}
\includegraphics[height=1.9in]{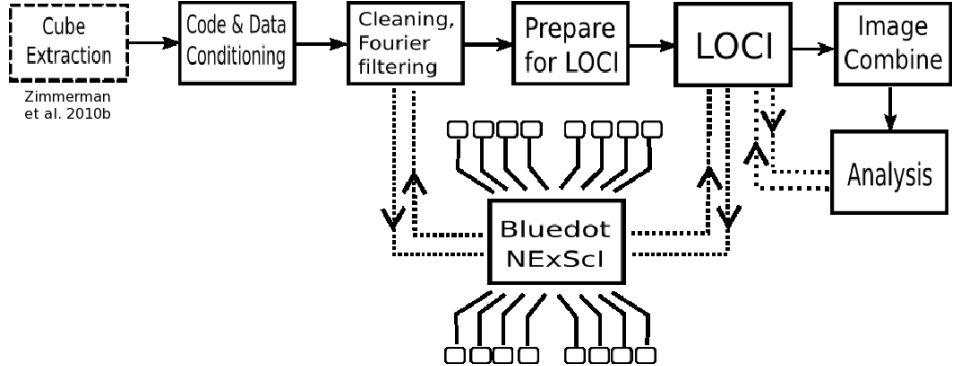}
\caption{Block diagram of the P1640 speckle suppression pipeline (PSSP). Images are first extracted from raw IFS data via the procedure described in Zimmerman et al. 2010b to form cubes. The cubes are then processed using wavelength diversity and the LOCI algorithm to detect faint companions that are originally buried in speckle noise. Multiple CPU nodes from the NASA Exoplanet Science Institute (NExScI) Bluedot super-computer run custom Matlab scripts in parallel to enhance speed at several stages.}
\end{center}\label{fig:block}
\end{figure*} 

Depending on target near-infrared brightness, several dozen cubes are generated for each star. This large data set, which we refer to as a hyper-cube when the time dimension is included (Fig. 3), consists of hundreds of images. In each channel of each cube, the spatial pixels are cleaned using standard techniques to correct for any remaining intensity outliers and negative values. Depending on whether we are searching for point sources or diffuse circumstellar material, like a dust disk, the images may be high-pass Fourier filtered to remove the low frequency component of the stellar halo and enhance the signal of both companions and speckles -- the latter of which is useful for image alignment. The low frequency component of the image is removed using a quadratic attenuation profile, which we find has better noise properties than a linear or step-function filter. This stage of the procedure is time-consuming and so is sent to the NExScI Bluedot cluster for processing. 

Once cleaned, the images in each wavelength channel are stacked across the hyper-cube time dimension to form composite images (Fig. 3). This step increases the companion signal-to-noise ratio in preparation for the LOCI algorithm. The speckles themselves are used to register images. We find that this approach is efficient and as reliable as using the astrometric grid spots for alignment, provided a custom gray-scale mask is used to prevent the coronagraphic spot or companion itself from biasing the result. A Fourier-based program registers the images with 0.01 pixel precision \citep{guizar_08}. We find that allowing for small distortions in the position of speckles, by treating local image regions as a flexible membrane, results in only a marginal improvement in PSF matching. The number of temporal cubes used to form composite images depends on data quality and quantity. In general, cubes with sufficiently high speckle signal-to-noise ratio may be divided into a larger number of groups that are treated separately by the PSSP. The results may then be combined following image subtraction. 

The LOCI algorithm constructs an optimal PSF reference for each wavelength channel of the composite cube using neighboring frames in color and time. To generate sufficient wavelength diversity, and therefore minimize subtraction of companion light, the reference frames associated with a given composite image must be separated by several spectral channels, $\Delta c=c_o-c_i$, where $c_o$ is the channel of the composite image with wavelength $\lambda=\lambda_0$ and $c_i$ is the channel common to a given set of reference frames. The minimum difference between channels is a function of composite image wavelength, $\lambda_0$, spectral resolution, $\delta \lambda$, and field angle, $\theta_0$, 
\begin{equation}
|\Delta c_{\mbox{\small{min}}}| \approx \frac{\alpha}{\theta_0} \frac{\lambda_0}{\delta \lambda},
\end{equation}
where $\alpha \approx 1$ is the fraction of a diffraction width that PSF reference images are radially stretched and $\theta_0$ is expressed in the same units (unitless). The optimal $\Delta c$ range is determined by the competing effects of companion throughput and speckle correlation between channels.  

Fig. 4 displays the degree of correlation between images as a function of wavelength and time following cleaning and optimal stretching and shrinking. All images are compared to a channel in the H-band corresponding to a wavelength of $\lambda_0=1.55\;\mu$m in the first cube. A box is drawn around the reference images that may be sent to the LOCI algorithm for this case. The difference between two sets of wavelength channels is labeled. We generally include channels in the water-bands, located around $1.35 \lesssim \lambda \lesssim 1.50 \;\mu$m, despite their comparatively lower signal due to atmospheric absorption. It is likewise possible to use longer wavelengths to achieve chromatic diversity. This figure demonstrates the utility of an IFS for speckle suppression. Speckles in stretched images are highly correlated in color -- the degree to which is limited only by the intrinsic wavelength dependence of the instrument, but decorrelate after several minutes in time. 

\begin{figure*}[!t]
\begin{center}
\includegraphics[height=1.6in]{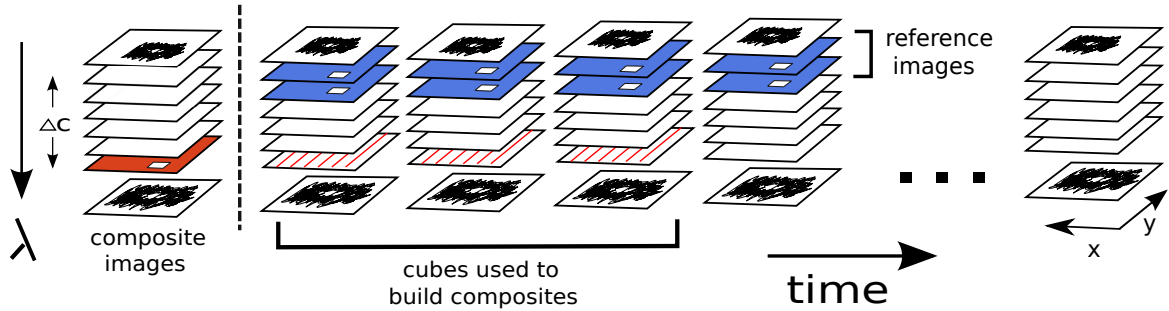}
\caption{Schematic of a P1640 hyper cube. Composite images shown to the left of the vertical dashed line are built by co-adding frames in each wavelength channel across the time domain to increase companion signal-to-noise ratio prior to performing PSF subtraction. In this example, images with red stripes are used to create the image in solid red. An optimal PSF reference image for each composite is constructed from the many available reference frames in the color and time domains (solid blue images). A small sub-frame is shown to indicate that PSF subtraction occurs locally. Composite images and usable reference frames are separated by $\Delta c$ wavelength channels to minimize subtraction of companion signal.}
\end{center}\label{fig:hyper_cube}
\end{figure*} 

Reference images that are closely spaced in time with $\Delta c>\Delta c_{\mbox{\small{min}}}$ are optimally stretched using an iterative least-squares technique for speckle spatial matching that incorporates precision registration via the Fourier-based method mentioned above. Images are resampled using bicubic interpolation, where the intensity of an output pixel is calculated from a weighted average of the neighboring 4x4 pixels. A first guess for the optimal stretching parameter, $S$, is based on the wavelength of channels of interest using the scaling relationship, $S=\lambda_0/\lambda$, expected from speckle movements for phase aberrations located solely in the pupil plane. Here $\lambda_0$ again represents the wavelength of a non-reference (composite) image and $\lambda$ represents the wavelength of a reference image. From this starting point, we then adjust the scaling factor to converge to an optimal value accurate to $\Delta S \approx10^{-3}S$. 

\begin{figure*}[!t]
\begin{center}
\includegraphics[height=2.12in]{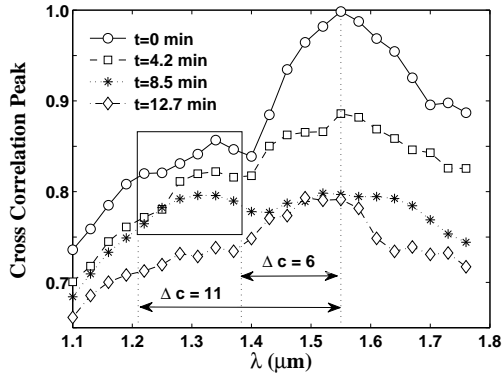}
\caption{Cross-correlation peaks in color and time for the star Alcor after optimal image stretching. The star is occulted by the coronagraph, the cubes have been high-pass Fourier filtered, and Alcor B has been masked. All images are compared to the $\lambda_0=1.55 \;\mu$m channel in the first cube. The legend shows the amount of time elapsed in minutes between exposures. Speckles are highly correlated as a function of wavelength (see also Fig. 1), thus justifying the use of processed IFS data for PSF reference subtraction. When constructing an optimal reference image via the LOCI algorithm, several channels must be skipped (given by $\Delta c$) to prevent subtraction of any companions. We nominally use cubes acquired within a certain interval of time. These images are identified as enclosed by a box. Different channels have different reference images. In this example, there are 18 highly correlated images available that provide sufficient wavelength diversity for the $1.55 \;\mu$m channel. More images may be used in general since the LOCI algorithm will assign their associated coefficients an appropriate (lower) weight.}
\end{center}\label{fig:cross_corr_peak}
\end{figure*} 

\begin{figure*}[!t]
\begin{center}
\includegraphics[height=2.12in]{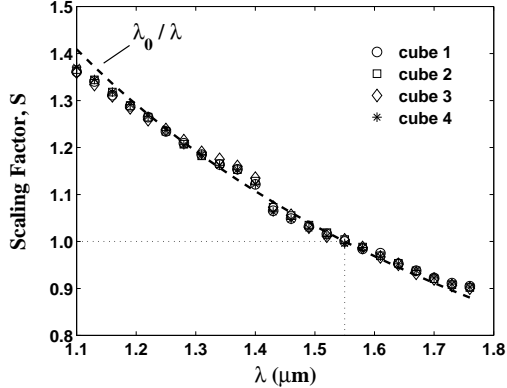}
\caption{Image scaling (stretching/shrinking) factors as a function of wavelength for all IFS channels calculated for $\lambda_0=1.55 \; \mu$m using several different cubes of the star Alcor. If all aberrations where phase errors that occurred in the pupil plane, the optimal scaling factors would follow $S=\lambda_0/\lambda$, shown by the dashed line. Since aberrations are also generated in amplitude and by out-of-pupil-plane optics in phase, $S<\lambda_0/\lambda$ when $\lambda<\lambda_0$, and $S>\lambda_0/\lambda$ when $\lambda>\lambda_0$ (see discussion). Image channels located between the J and H filters, near $\lambda\approx1.35-1.50 \;\mu$m, have a lower signal as they are affected by atmospheric water-bands, leading to a systematic offset and larger dispersion in the values by which images are scaled as a consequence of the cube extraction algorithm. We nominally use channels located in the water-bands to build PSF reference images for bright stars. Optimal scaling factors are therefore calculated on an individual basis.}
\end{center}\label{fig:scaling_factor}
\end{figure*} 

\begin{figure*}[!t]
\begin{center}
\vspace{0.3in}
\includegraphics[height=2.5in]{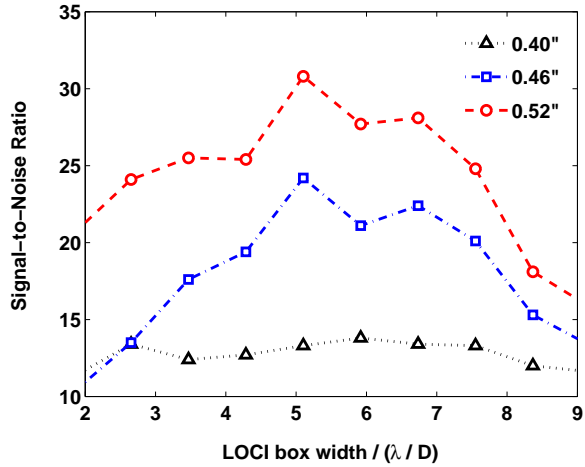}
\caption{Signal-to-noise ratio versus LOCI (outer) box width for three artificial companions injected into the Alcor data set shown in Fig. 7. The optimal LOCI box has a width of $\approx5\;\lambda/D$, reaching a compromise between speckle suppression and companion throughput. The companions have equal brightness and separations of $0.40\arcsec, 0.46\arcsec,$ and $0.52\arcsec$. More wavelength diversity ($\Delta c>6$) is required to detect companions with separation $<0.4\arcsec$.}
\end{center}\label{fig:loci_box}
\end{figure*} 

\begin{figure*}[!t]
\begin{center}
\vspace{0.2in}
\includegraphics[height=2.5in]{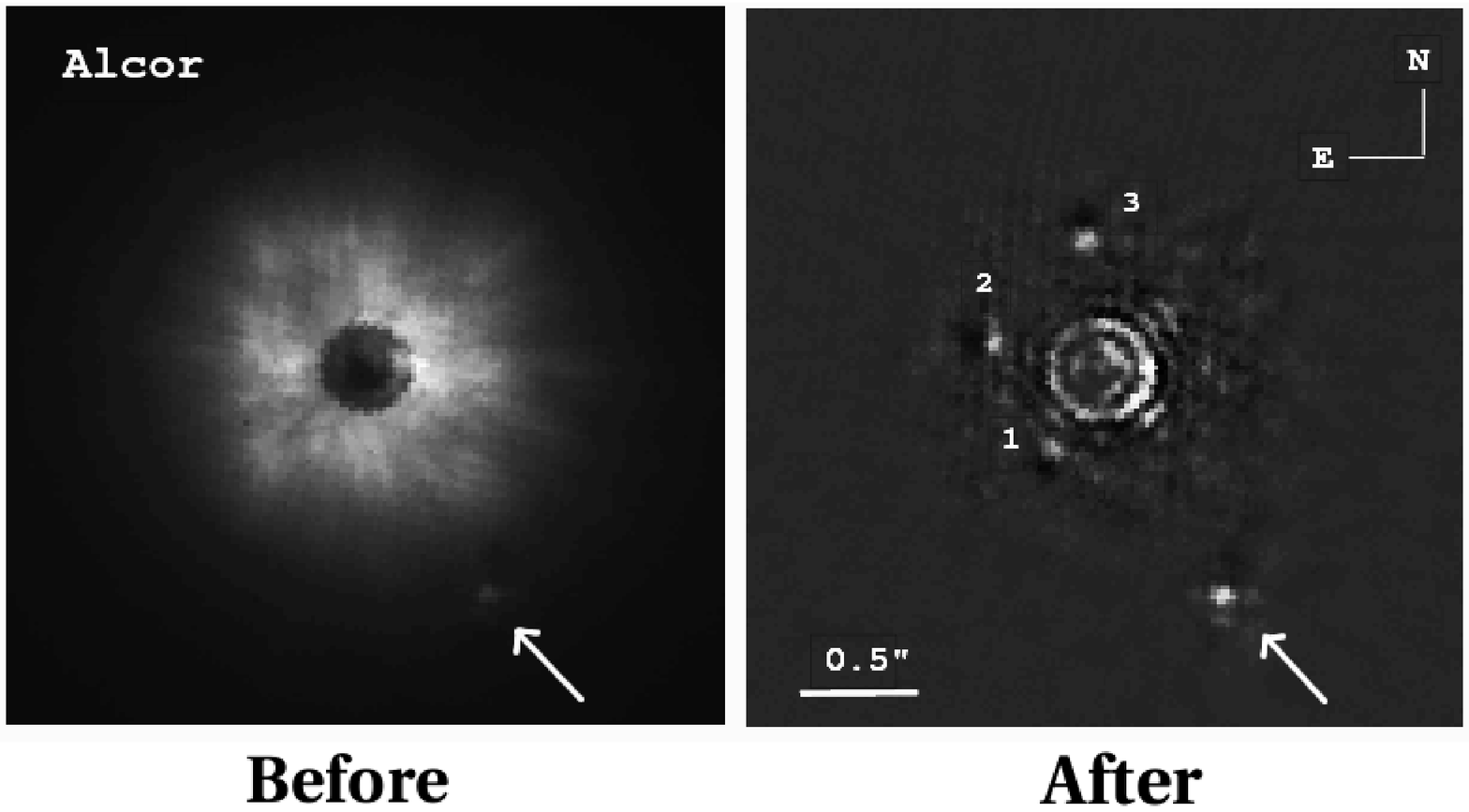}
\caption{Stacked H-band images of the star Alcor before (left) and after (right) running the PSSP. Three artificial companions have been injected into the on-sky data with separations of $0.40\arcsec, 0.46\arcsec$, and $0.52\arcsec$. The companions are 1 magnitude fainter than Alcor B in each channel and detectable only after using the wavelength diversity provided by the IFS. The dipole pattern is evident for the inner-companions as a consequence of image stretching. Alcor B does not exhibit the dipole pattern because it is masked in PSF reference images to facilitate image registration.}
\end{center}\label{fig:alcor}
\end{figure*} 

Figure 5 shows the optimal scaling factors, $S$, plotted as a function of wavelength for the same reference channel and dataset shown in Fig. 4. The theoretical relationship, $S=\lambda_0/\lambda$, is over-plotted for comparison. With each cube, we find that optimal scaling factors are systematically smaller than the expected $\lambda_0/\lambda$ relation when $\lambda<\lambda_0$, but larger than $\lambda_0/\lambda$ when $\lambda>\lambda_0$. This effect is understood intuitively by considering the radial displacement of speckles as a function of wavelength. The scaling relation, $S=\lambda_0/\lambda$, is derived for phase errors located in a pupil plane. In practice, wave-front phase errors are also generated by out-of-pupil-plane optics and reflectivity and transmission non-uniformities which can transform into phase aberrations \citep{shaklan_green_06}. Consequently, the radial position of a given speckle, $\theta(\lambda)$, increments more slowly than the linear dependence expected from first-order diffraction theory and instead follows $\theta(\lambda) \lesssim \lambda \; \theta_0$ as a function of wavelength. 

Exceptions occur in the water-bands ($\lambda \approx 1.4 \;\mu$m) as a consequence of the cube extraction algorithm operating with low signal. In this regime, the spectrum in each spatial pixel becomes biased towards the red-end of the J-band and also blue-end of the H-band (see Zimmerman et al. 2010b for details). As a result, the cube extraction algorithm creates a spatial pattern that systematically mimics an adjacent spectral channel, thus creating the bump seen in Fig. 5. This systematic error does not effect PSF subtraction, but it does artificially modify the spectrum of companions in the water-bands, requiring calibration. The rms value of $S$ averaged across the spectral bands is 0.003, comparable to image alignment precision. As a result of these effects, the PSSP recalculates the optimal scaling factor for each image in each cube for each star to maximize performance. 	

Our implementation of the LOCI algorithm operates on each spatial pixel in turn. This step is likewise time-consuming and so is also sent to the Bluedot cluster. Local image coefficients are generally assigned based on the flux inside of a box of width $5\times5$ PSF widths. Fig. 6 shows optimization of the signal-to-noise ratio as a function of the LOCI box width (referred to as the O-zone in \cite{lafreniere_loci}) for three artificial companions inserted into the Alcor data set. Their angular separation from the star are $0.40\arcsec, 0.46\arcsec, 0.52\arcsec$ respectively as shown in Fig. 7 and discussed in more detail below. Once complete, the program co-adds the reduced data in the color dimension. The J and H-bands are separated at $\lambda=1.40 \; \mu$m to avoid unnecessary averaging in case a companion is particularly red or blue. 


\section{On-Sky Demonstration}\label{sec:results}
P1640 has been used on several observing runs at Palomar in recent semesters. In this section, we present our preliminary on-sky results, demonstrating the ability to recover real and artificial companions that are faint compared to the intensity of quasi-static speckles. We characterize the signal-to-noise ratio of detections and show our sensitivity in terms of contrast. 

\subsection{Alcor}\label{sec:alcor}
P1640 discovered an M-dwarf orbiting the nearby, bright A-star Alcor in March 2009 \citep{zimmerman_10}. With an angular separation of $\approx$1$\arcsec$ and brightness ratio of 6 magnitudes, the $\approx0.25M_{\odot}$ companion was identified in stacked cubes prior to speckle suppression.\footnote{Extracted IFS data cubes may be converted into a movie that sequentially displays images in wavelength space, allowing for the visual identification of candidate companions whose brightness is comparable to the intensity of speckles.} In this section, we use the Alcor data set to demonstrate the ability to detect artificial companions that are fainter and located closer to the primary star where quasi-static speckles are bright as a result of non-common-path wave-front errors.

To test the PSSP and also calculations of contrast ($\S\ref{sec:contrast}$), we inserted artificial companions of various brightness and angular separation into the Alcor on-sky data. The companions were given PSF widths and relative intensities that match Alcor B in each channel. To minimize PSF smearing in color and time, each companion was injected into each image channel of each cube by using Alcor B as an astrometric reference point. An optimal LOCI (outer) box size of $5\times5$ PSF widths was used to maximize the signal-to-noise ratio (Fig. 6). Fig. 7 shows H-band images of the star before and after application of the PSSP using three artificial companions that are each 1 magnitude fainter than Alcor B with separations of $0.40\arcsec$, $0.46\arcsec$, and $0.52\arcsec$ respectively.

The three artificial companions are completely buried in speckle noise in pre-processed cubes and detectable only after using the wavelength diversity provided by the IFS. After application of PSSP, each display the tell-tale dipole pattern -- a closely spaced positive and negative image -- as a result of reference image stretching and subtraction.\footnote{The inclusion of both image stretching and shrinking would yield a tri-pole pattern \citep{sparks_ford_02}. It is possible to further build companion flux by subtracting composite images from reference images, and then rescaling and stacking the results. This too creates a tripole pattern, but it also smears the PSF and can degrade astrometric precision.} Alcor B does not exhibit the dipole pattern because it is masked in each PSF reference image to facilitate precision image registration. The artificial companions have a final signal-to-noise ratio of 13, 24, and 31 for respective separations of $\theta=$ $0.40\arcsec$, $0.46\arcsec$, and $0.52\arcsec$ as shown Fig. 6. Alcor B is unambiguously recovered with a signal-to-noise ratio gain from 11.9 to 113.4. 
 
The artificial companions have a PSSP throughput of 42\% (inner), 35\% (middle), and 51\% (outer), where throughput is defined as the ratio of the total companion flux before and after data processing averaged over each wavelength channel in the H-band. Throughput is a function of the local speckle noise and angular separation and must be calibrated to accurately recover the spectro-photometric signal of companions that are fainter than stellar artifacts prior to applying PSSP. We find that broadband aperture photometry measurements typically result in $10\%$ errors following calibration, with larger uncertainties occurring closer to the star. More precise measurements may be obtained by selecting PSF reference images that minimize fluctuations in the speckle subtraction residuals, but even $5\%$ photometry is challenging given the many inherent biases present in LOCI datasets that must be accounted for (e.g., \cite{marois_10spie}). The topic of precise spectral extraction is discussed in detail in Pueyo et al. 2011. 

The bright ring surrounding the coronagraph occulting spot results from insufficient wavelength diversity as a consequence of the slower outward movement of speckles with spectral channel in the regions very close to the star. This signature may be used to qualitatively assess the quality of reference image alignment. Residual starlight is distributed symmetrically, but is more intense on one side of the coronagraph as a result of tip/tilt errors experienced in several data cubes. 

No real companions closer to the Alcor primary were detected. We find that artificial companions injected with a yet fainter signal, by factors of $\approx$2.6, 4.8, and 6.2 at separations of $\theta=0.40\arcsec, 0.46\arcsec$, and $0.52\arcsec$, are recoverable respectively. This result is consistent with Fig. 6 and was also used to check PSSP contrast calculations (see $\S$\ref{sec:contrast}). Alcor is a member of the Ursa Majoris moving group which has a purported age of 400-600 Myrs \citep{castellani_02,king_03}. Using the \cite{zimmerman_10} H-band photometry of the primary and secondary along with our contrast measurements, we are able to rule out additional companions with mass $m\geq77M_J$ and separation $\theta \geq 0.52\arcsec$ \citep{baraffe_03}. 

\subsection{FU Orionis}\label{sec:fuori}
FU Orionis (Ori) is a prototypical young stellar object after which a class of enigmatic low-mass stars exhibiting signs of disk accretion is named. Such objects experience outbursts resulting in 4-5 magnitudes of visual brightening on time scales of months with decay over decades sometimes punctuated by smaller scale episodes \citep{hartmann_kenyon_96}. FU Ori itself has a known stellar companion separated by $0.5\arcsec$ \citep{wang_04}. It is thought that many FU Ori objects are members of binary or hierarchical triple systems \citep{reipurth_aspin_04}. 

We observed FU Ori on March 17, 2009 to better characterize the companion which we measure to be 5.2 magnitudes fainter than the primary in the J-band and 4.4 mags fainter in the H-band. Given the level of contamination, FU Ori B does not currently have a published J-band spectrum as it is mixed with speckle noise. Indeed, it was originally discovered using PSF subtraction of a nearby calibration star \citep{wang_04}.

\begin{figure*}[!t]
\begin{center}
\includegraphics[height=2.4in]{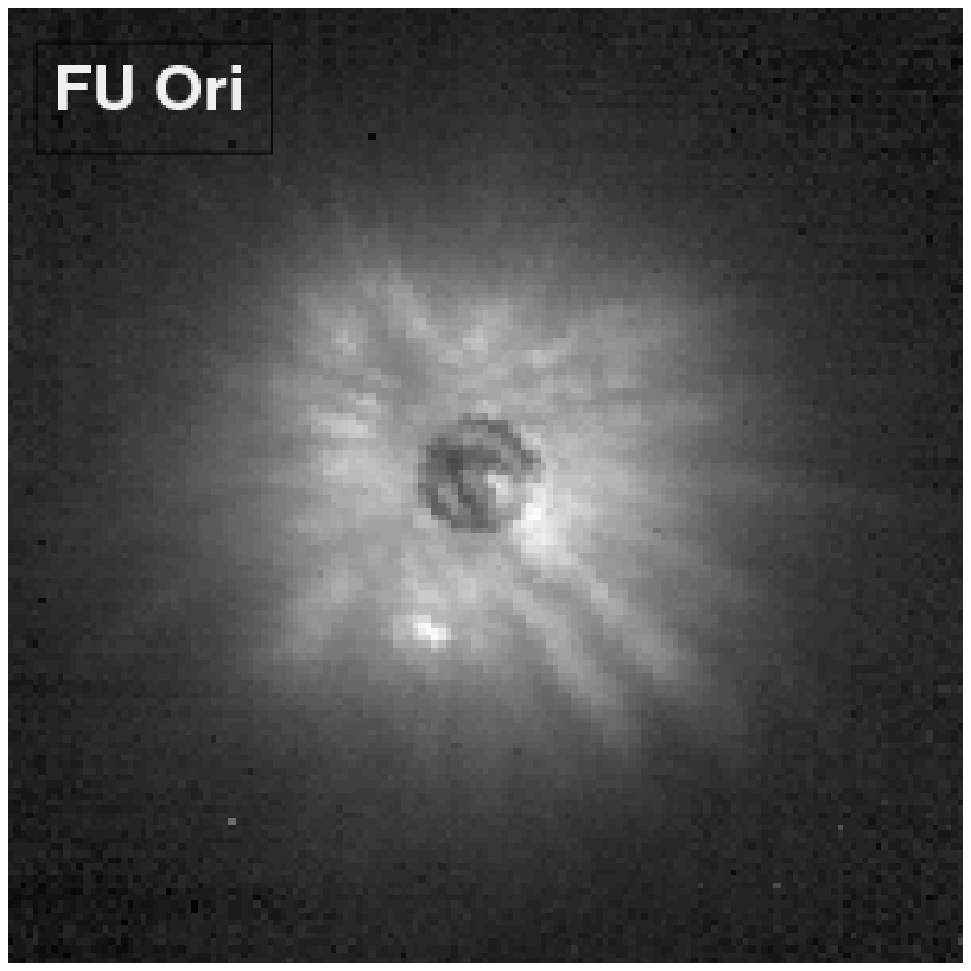} 
\includegraphics[height=2.4in]{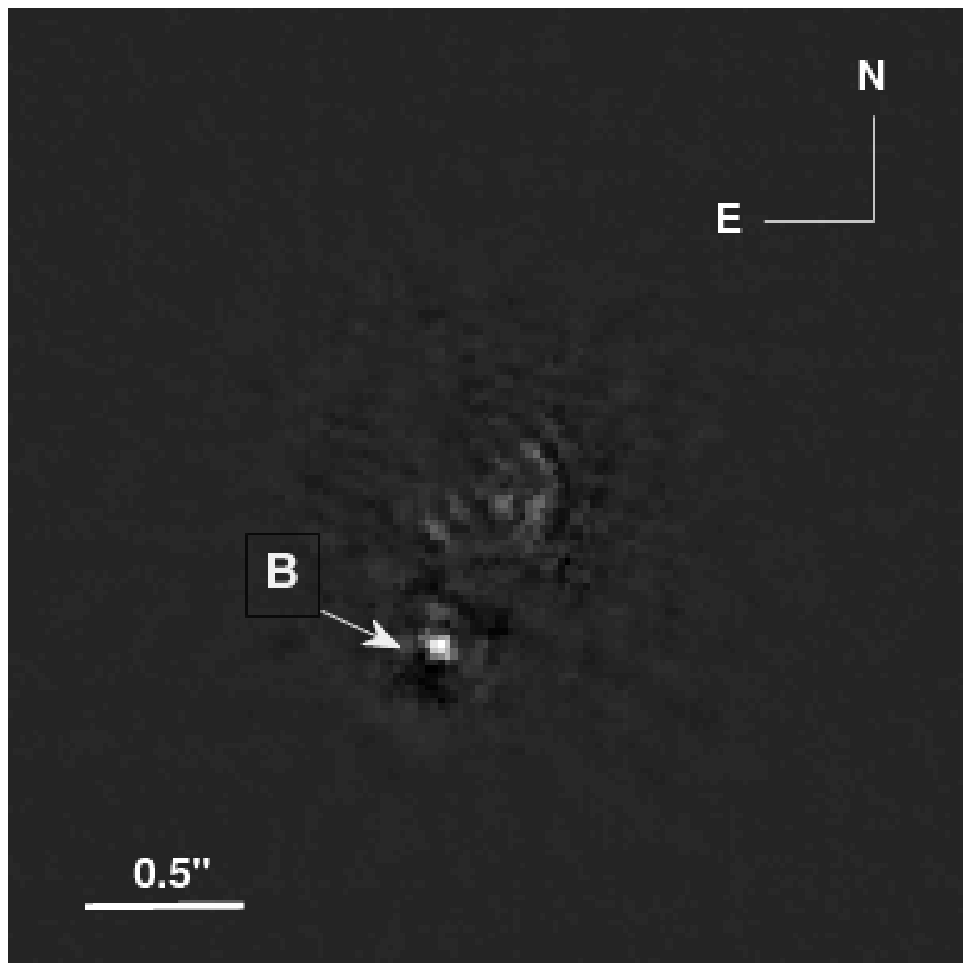}
\caption{Stacked JH images of the star FU Ori before (left) and after (right) running the PSSP. The companion, FU ORI B, is located within the AO control box and has a brightness comparable to the scattered-light halo of the primary. Its unambiguous detection following post-processing demonstrates the principal of using an IFS to discriminate between speckles and faint off-axis sources.}
\end{center}\label{fig:fu_ori}
\end{figure*} 

Fig. 8 shows combined JH images of FU Ori before and after running the PSSP algorithm. The companion is just perceptible in pre-processed data. Once applied, the PSSP unambiguously recovers the companion as indicated by the bright positive and negative signature. Analysis of individual channels shows that we are able to improve from a 1.5$\sigma$ to 16.4$\sigma$ detection at $\lambda=1.22 \;\mu$m and 2.3$\sigma$ to 22.9$\sigma$ detection at $\lambda=1.58\;\mu$m, where $\sigma$ is the local standard deviation in speckle noise intensity. The J and H-band spectrum of FU Ori B will be presented in a subsequent paper. This experiment explicitly demonstrates the utility of using an IFS for speckle-suppression applications. 

\subsection{Measured Contrast}\label{sec:contrast}
The PSSP provides sensitivity measurements for each star reduced. Contrast is calculated at several stages during the reduction, including with raw (pre-processed) data, high-pass Fourier-filtered data, and speckle suppressed data. Speckle noise is found by measuring the local standard deviation of intensity variations occurring in a box of size 5$\times$5 PSF widths, where flux levels are calculated on the scale of a single PSF by averaging over several pixels. Speckle noise values are then compared to the stellar peak intensity measured in unocculted frames to calculate the relative brightness of directly detectable companions. We also estimate the noise floor set by photon arrival statistics by measuring the square-root of the mean flux over the same regions. We nominally quote 5$\sigma$ contrast levels which correspond to a signal-to-noise ratio where companions become noticeable by eye.\footnote{Speckle noise follows a Rician spatial distribution and speckles with intensity up to $\approx$10$\sigma$ may occur \citep{aime_soummer_04,fitzgerald_graham_06}. An IFS however can help to discriminate between unusually bright speckles and real companions by comparing the spectrum of the candidate to that of the star.}

Results for the $V=6.7$ star HD 204277 from data taken on June 28, 2009 are shown in Fig. 9. Local contrast levels using raw, Fourier-filtered, and speckle-suppressed images have been azimuthally averaged (median). We find that residual quasi-static wave-front phase errors of $\approx$100 nm rms limit raw (pre-processed) sensitivity to $10^{-3}$ at the coronagraph inner-working-angle. This result is consistent with similar measurements using the PHARO instrument prior to application of the modified Gerchberg-Saxton phase retrieval algorithm \citep{burruss_10}. High-pass filtering the data removes the low-frequency component of the stellar halo (pedestal) resulting in a factor of several improvement, most notably at separations exterior to the AO control region at $\approx$$0.6\arcsec$. Finally, chromatic differential imaging operating in tandem with the LOCI algorithm reduces remaining noise by an order of magnitude closer to the star. In the case of HD 204277, this results in a $5\sigma$ H-band contrast of $2.1\times10^{-5}$ at $1\arcsec$ with 20 minutes of on-source integration time. 

\begin{figure*}[!t]
\begin{center}
\includegraphics[height=2.5in]{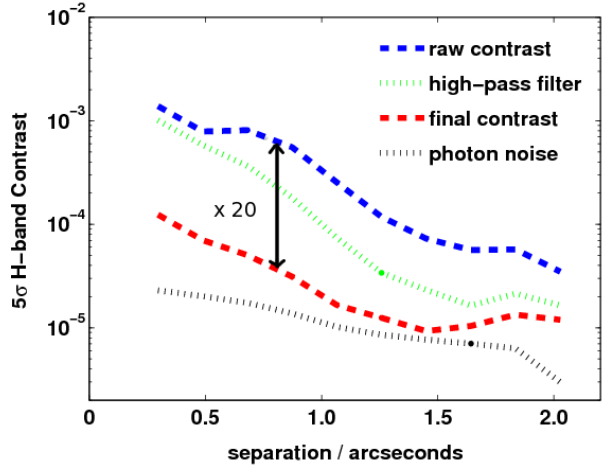}
\caption{P1640 contrast levels in the H-band for the star HD 204277 taken on June 28, 2009 UT. The star is occulted by the apodized pupil Lyot coronagraph. PSSP consistently provides a gain of at least 1 order of magnitude in sensitivity at subarcsecond separations.}
\end{center}\label{fig:contrast}
\end{figure*} 

These are the deepest contrast levels yet achieved at Palomar in the H-band. Sensitivity is currently limited by calibration of non-common-path wave-front errors and instrument transmission which ultimately governs the level of photon-noise and signal-to-noise ratio of speckles. We find that contrast continues to improve with additional exposures, but slower than a square-root relationship in time in regions very close to the star. 
	
\section{Summary \& Concluding Remarks}\label{sec:summary}
Central to the issue of generating contrast levels sufficient to directly image extrasolar planets is subtraction of (unavoidable) quasi-static speckles that arise from instrument optical aberrations. Along with the ability to characterize companions, an integral field spectrograph naturally provides the capability for suppressing this dominant noise source. The recently commissioned Project 1640 instrument at Palomar is the first to use an IFS for high-contrast imaging in combination with a coronagraph and adaptive optics. 

We have written a custom program, the Project 1640 speckle suppression pipeline (PSSP), to improve instrument raw sensitivity via post-processing. The program utilizes a version of chromatic differential imaging, or spectral deconvolution, to perform point-spread-function subtraction, as was first proposed by Sparks  \& Ford 2002. To this, we also use the multitude of images provided by the IFS to construct PSF references via the locally optimized combination of images (LOCI) algorithm \citep{lafreniere_loci}.  

We have quantified the degree of spatial correlation between speckle patterns in adjacent P1640 hyper-cube images as a function of color and time. We find that spectral image diversity provides PSF references that match science images at a level comparable to, if not better than, observations of a nearby calibration star, depending on over-head time and system flexure. Analysis related to image scaling factors, throughput, and signal-to-noise ratio well-match theoretical expectations and results from other speckle diversity techniques that also use LOCI, once relevant parameters are optimized. Further, we are able to perform pseudo-real-time data processing while at the observatory using the Bluedot super-computing cluster at NExScI,

We have quantified the improvement afforded by the PSSP with three different data sets. Injecting faint companions, recovering known companions, and calculating contrast levels before and after performing speckle suppression each indicate that on-sky sensitivities consistently improve by at least one order of magnitude at angular separations within the AO control region following post-processing. At larger separations, the intensity of residual starlight approaches the limit set by photon noise. Our sensitivity currently reaches $5\sigma$ contrast levels of $\approx2.1\times10^{-5}$ at $1\arcsec$ in the H-band with 20 minutes of integration time. This result is comparable to other experiments at Palomar and is currently limited by calibration of non-common-path errors \citep{burruss_10,crepp_10}. Preliminary tests using an interferometric wave-front calibration unit show promise to reduce these errors by an order of magnitude when operating in tandem with the PALM-3000 ``extreme" AO system, suggesting that contrast levels of order $\approx10^{-7}$ inside of an arcsecond are feasible with integration times of 1-2 hours (Shao et al. 2011, in prep., \cite{bouchez_09}). 

IFS data provides the ability to play movies by sequentially stepping through each image in the color dimension. We have performed experiments carefully comparing our ability to detect faint companions using these color-movies and the PSSP. We find that movies perform remarkably well, and that the PSSP generally provides an additional factor of 2-3 in effective contrast in comparison. This gain is most noticeable in close proximity to the star, just exterior to the coronagraphic spot, and is largely a result of implementing the LOCI algorithm to improve PSF matching. We play color-movies as part of our standard reduction package to identify companions with brightness comparable to raw speckle noise. Stepping through wavelength channels is likewise helpful with differenced images.

It is necessary to calibrate the effects of partial subtraction resulting from LOCI to accurately measure the relative intensity of a faint companion \citep{marois_08,thalmann_09,currie_10,bowler_10,janson_10gj758,marois_10spie}. This is nominally accomplished by injecting artificial companions with known brightness into each data cube. Our photometric analysis indicates that $10\%$ uncertainties are typical for angular separations within the AO control region and that larger uncertainties, by factors of several, may occur very close to the star where only marginally sufficient wavelength diversity is achieved. 

High-contrast astrometry is likewise challenging. The inclusion of astrometric grid spots aids with locating the position of the occulted star, but aggressive differential imaging techniques modify the shape and flux distribution of the companion PSF, degrading precision. Our preliminary analysis using artificial companions indicates that systematic errors of size $\approx0.5$ spatial pixels are common, even when using companion masks in reference frames. Carefully chosen reference frames can help to mitigate these effects, but 1 mas astrometry using near-Nyquist sampled data ($\sim$2-4 spatial pixels per diffraction width in this case) may prove to be prohibitive unless more advanced algorithms are developed. 

This paper demonstrates one of the two important benefits provided by an IFS for high-contrast imaging: the automatic generation of useful PSF references for chromatic differential imaging and the detection of faint companions. A subsequent paper by Pueyo et al. 2011 will discuss the challenges and possible solutions for accurately extracting their spectra.

\section{Acknowledgements}
We are grateful to the staff at Palomar Observatory for their support. Project 1640 is funded by National Science Foundation grants AST-0520822, AST-0804417, and AST-0908484. LP and SH acknowledge support from the Carl Sagan Fellowship Program. A portion of the research presented in this paper was carried out at the Jet Propulsion Laboratory, California Institute of Technology, under contract with the National Aeronautics and Space Administration (NASA). 

\begin{small}
\bibliographystyle{jtb}
\bibliography{ms.bib}
\end{small}

\end{document}